\documentclass[prl,showpacs,floatfix,twocolumn]{revtex4}
\usepackage{graphicx}
\newcommand{\be}{\begin{eqnarray}}
\newcommand{\ee}{\end{eqnarray}}
\def\beq{\begin{equation}}
\def\eeq{\end{equation}}

\begin{document}
\title{A semiclassical theory of the Anderson transition}
\author{Antonio M. Garc\'{\i}a-Garc\'{\i}a}
\affiliation{Physics Department, Princeton University, Princeton,
New Jersey 08544, USA}
\begin{abstract}
We study analytically 
the metal-insulator transition in a disordered conductor by combining the self-consistent theory of localization with the 
one parameter scaling theory.  We 
 provide explicit expressions of the critical exponents and the critical disorder as a function of the spatial dimensionality $d$. 
The critical exponent $\nu$ controlling the divergence of the localization length at the transition is found to be
$\nu = {1 \over 2 }+ {1 \over {d-2}}$ thus confirming that the upper critical dimension is infinity. 
Level statistics are investigated in detail.  We show that the two level correlation function decays exponentially and the number variance is linear with a  
slope which is an increasing function of the spatial dimensionality.   Our analytical findings are in agreement with previous numerical results.
\end{abstract}

\maketitle
 The recent experimental realization of disorder in ultra cold atoms \cite{exp} together with the rapid progress in numerical calculations \cite{numand,sko,schreiber,sch1,ever,antem2} has revived 
 the interest in the metal-insulator transition (MIT) \cite{anderson}.
%For non interacting, time reversal systems with short range disorder it was found $s  \approx 1, \nu \approx 1.5$  
%where $s,\nu$ stand for the critical exponents related to the vanishing of the conductivity $\sigma$ and the divergence of the localization length $\xi$ respectively.
The statistical analysis of the spectrum and eigenvectors plays a central role in the identification and characterization of a MIT.
Typical signatures of a MIT include: a) multifractal \cite{ever,schreiber} eigenstates
(for a review see \cite{mirlin,cuevas}), namely, the scaling of
${\cal P}_q=\int d^dr |\psi({\bf r})|^{2q} \propto L^{-D_q(q-1)}$
with respect to the sample size $L$ is anomalous with $D_q < d$ a set of exponents describing the
transition, b) scale invariance \cite{sko} of the spectral correlations, 
c)  level repulsion of neighboring eigenvalues as in a disordered metal, d) linear number variance like in a disordered insulator, 
 $\Sigma^2(\ell)=\langle (N_\ell -\langle N_\ell \rangle)^2 \rangle \sim \chi \ell$  
($N_\ell$ is the number of eigenvalues in an interval of length $\ell \gg 1$ and $\langle \ldots \rangle$ stands for ensemble average) but with a slope $\chi <1$. 
In 3$d$, $\chi \approx 0.27$ for higher dimensions see Ref.\cite{antem2,sch1}. 
Level statistics with these features are
usually referred to as critical statistics \cite{sko,KMN,chi}.  
 
Theoretical progress has been much slower in recent times. 
With the exception of the case of a Cayley tree \cite{MF} geometry we are indeed far from a quantitative analytical theory
of the MIT. Below we briefly review the main analytical approaches to the MIT problem.   
In the original Anderson's paper the critical disorder at which the MIT occurs was estimated  by looking at the
limits of applicability of a locator expansion \cite{anderson,thouless}.  In this approach \cite{anderson} the metal-insulator
transition is induced by increasing the hopping amplitude of an initially localized
particle. A more
refined self-consistent \cite{abu} condition, still within the locator formalism, 
provided with a similar answer.  
This method is only exact in the case of
a Cayley tree but it is believed to be accurate in the $d \to \infty$ limit.  
In both cases a MIT in 3$d$ is predicted correctly. However the estimated critical disorder is considerably smaller than the one found 
in numerical simulations \cite{schreiber}. This disagreement persists in higher dimensions \cite{antem2}.

The one parameter scaling theory \cite{one} (OPT) provides a completely different approach to the MIT.  
A key concept in this theory is the dimensionless conductance $g= E_c/\Delta$ \cite{thouless2} where, $E_c$, the
Thouless energy, is an energy scale related to the diffusion time
to cross the sample and $\Delta$ is the mean level spacing. In a metal, $E_c = \hbar
D_{clas}/L^2$ ($D_{clas} = v_F l /d$ is the classical diffusion constant with $l$ the mean free path and $v_F$ the velocity of the particle) and
therefore $g \propto L^{d-2}$. In an insulator the particle
is exponentially localized and $g
\propto e^{-L/\xi}$ where $\xi$ is the localization length.  The OPT is based on the following two simple
assumptions:  a) $\beta(g)=
\frac{\partial \log g(L)}{\partial \log L}$ is continuous and
monotonous, b) The change in the conductance with the system size
only depends on the conductance itself. 
With this input the OPT predicts correctly a MIT for $d>2$  characterized
by a size independent dimensionless conductance $g = g_c$ such that
$\beta(g_c) = 0$ and $\beta(g) > (<)~0$ for $g > (<)~g_c$. 
In order to make more quantitative statements about the MIT it is necessary to understand  in detail
how the system approaches the transition.  

In the case of  $d = 2 + \epsilon$ ($\epsilon \ll 1$) the MIT occurs at weak disorder. A rigorous analytical treatment is possible by combining 
diagrammatic perturbation 
theory and field theory techniques \cite{wegner,hikami,voller}.  The results thus obtained for critical exponents and critical disorder 
are in agreement with numerical calculations \cite{mirlin}.
Later on Vollhardt and Wolfle \cite{voller1} proposed an extension of this theory valid for any $d$.  The idea was to go 
beyond perturbation theory around the metallic limit by  
solving a self-consistent equation for a renormalized diffusion coefficient.  The self-consistent condition was obtained 
by relating 
ladder diagrams associated with difussons with crossed diagrams associated with cooperons.  Time reversal invariance is a required condition for this relation to hold.    
As in the $d =2+\epsilon$ case, the MIT is induced by the growing effect of destructive interference on an otherwise metallic state.  
However a similar self consistent condition was also obtained starting from a locator expansion \cite{kroha}. Unfortunately some of the results of the Vollhardt-Wolfle theory 
do not agree with recent numerical results: for instance it is predicted that the 
upper critical dimension for localization is $d=4$ and that the critical exponent, $\nu$, controlling the divergence of the 
localization length is  $\nu =1/(d-2)$. However numerical results indicate the upper critical dimension for localization is $d > 6$
and $\nu  \approx 1.5, 1$ in 3$d$, 4$d$ respectively \cite{antem2,sch1}.

From the above discussion it seems clear that except in the $d =2 +\epsilon$  limit the MIT cannot be described by any perturbation theory around the metallic or the insulator 
side. The ultimate reason for that it is that, according to the OPT,  the MIT is a fixed point, $\beta(g_c) =0$, of the motion.   
It is a well known fact that in these situations the Hamiltonian may have universal properties completely different from the ones observed in the proximity of the transition. The above theoretical approaches neither can be extrapolated to the physically relevant case of $d =3$ nor predict the dependence with the spatial dimensionality of different parameters describing the transition.

In order to make progress in this problem a deeper knowledge of the dynamics precisely at the transition is needed.
This paper is a first step 
in that direction.  We solve the 
selfconsistent condition of Vollhardt-Wolfle \cite{voller,voller1} but including the spatial dependence of the diffusion constant predicted by the OPT.  
With this simple input we
obtain explicit expressions of different parameters characterizing the MIT such as critical exponents and the slope of the number variance as a function of the spatial dimensionality $d$.  The analytical expressions for critical exponents and level statistics are in agreement with previous numerical calculations \cite{sch1,antem2,ever}. 
 Throughout the paper we assume no interaction, time reversal invariance and periodic boundary conditions. 
The results of this letter are thus not applicable to the  MIT of the integer quantum hall effect \cite{qef} or in the one occurring in a 
2d disordered systems with spin-orbit interactions \cite{spt}. The  extension of this semiclassical formalism to these cases will be published elsewhere \cite{ant30}.

{\it One parameter scaling theory and anomalous diffusion at the MIT.-}
In the metallic limit, $g \to \infty$, the dynamics of a single particle in a random potential 
is well described by a normal diffusion process. The density of probability $P(\vec r,t)$ of finding a particle, initially at the origin, around the position $\vec r$ at time $t$ is described by the solution of the diffusion equation, 
$P(\vec r,t)   = \frac{e^{-|{\vec r}|^2/D_{clas}t}}{{(2D_{clas}t)}^{d/2}}$ in real space and  
$P(\omega,q) = \frac{1}{-i\omega + D_{clas}q^2}$ in Fourier space
where $|\vec q|^2 \equiv q^2$.
Since $\langle r^2 \rangle = D_{clas}t$, $g \propto L^{d-2} \gg 1$ for $d > 2$. 
These expressions are the starting point to study transport properties and level statistics in the metallic limit or when the localization corrections are small \cite{mirlin}.
The MIT for $d \geq 3$ occurs in the strong disorder region which is beyond the range of applicability of perturbation theory. However the
OPT can predict the type of motion at the critical point for any dimension: at the MIT 
$g = E_c/\Delta$ does not depend on the system size.  Since $\Delta \propto 1/L^d$,  the Thouless energy must scale as $E_c \propto 1/L^d$. 
This only can happen if the diffusion at the MIT is anomalous with,
$\langle r^{2m}\rangle \sim t^{2m/d}$ \cite{anoma}
where $m$ is a positive integer.  This result can likewise be interpreted as that the diffusion constant become 
scale dependent $D(L) \propto 1/L^{d-2}$ or, in momentum space, $D(q) \propto q^{d-2}$.
The OPT is only capable to predict $\langle r^2 \rangle$ but not the 
distribution function $P(\omega,q)$. However it is precisely this function the one needed for the evaluation of critical exponents or
level statistics. On the other hand it is evident that any perturbation theory around the metallic or insulator limit will fail if it cannot take into account the 
anomalous diffusion predicted by the OPT.  In fact anomalous diffusion in low dimensional systems has been related to a power-law decay of the eigenstates \cite{ant5,ever}. This strongly suggests \cite{ant30}that a new basis for the localization problem given by eigenstates with a power-law decay is the starting point for a meaningful perturbation theory at the MIT.

In order to proceed we have to come up with an expression for $P(\omega,q)$ consistent 
with the OPT prediction,  $\langle r^2 \rangle \sim t^{2/d}$ and that at the same time it can describe the dynamics in the proximity of the MIT. The simplest alternative is to 
assume that the classical diffusion coefficient $D_{clas}$ gets renormalized to $D = {\tilde D} (\omega){\tilde D}(q)$ with ${\tilde D}(q) =D_0 q^{d-2}$.  
The function ${\tilde D}(\omega)$ is given by the solution of the self consistent condition of Vollhardt-Wolfle,
%The self consistent equation for the diffusion coefficient including the momentum dependence term,
$
\frac{\tilde D(\omega)}{D_{clas}}=1-\frac{\Delta}{\pi \hbar V D_{clas}}\sum_q \frac{1}{- \frac{i\omega}{{\tilde D}(\omega){\tilde D}(q)} +q^2}.
$
%The range of diffusion coefficient is in principle restricted to the region of $q,\omega$ such that effects related to eigenstates 
%multifractality are negligible. This typically corresponds to long times but still shorter than the Heisenberg time $t_H \sim \hbar/\Delta$ and distances much larger than the mean free %path, $l$,  but smaller than the system size $L$.

As a first step to solve this self-consistent equation
we replace the sum by an integral and write down $\Delta$ and $D_{clas}$ as a function of $k_F = mv_F/\hbar$ and $l$,
$
\frac{\tilde D(\omega)}{D_{clas}}=1 - \frac{d k_F^{2-d}}{\pi k_F l}\int_0^{1/l}dq \frac{|q|^{d-1}}{\frac{-i\omega}
{{\tilde D}(\omega)D(q)}+q^2},
$
using $\frac{x^{2d-3}}{a^d + x^d} = a^d x^{d-3}\left[\frac{1}{a^d}-\frac{1}{a^d+x^d}\right]$ and noting that the effective disorder strength is  
 controlled by the parameter $\lambda \equiv 1/k_F l$  (the metallic limit corresponds thus with $\lambda \to 0$),
\be
\label{main}
\frac{\tilde D(\omega)}{D_{clas}} = 1-\frac{d}{(k_F l)^{d-1}(d-2)\pi} + \\ \nonumber
\frac{1}{D_0\xi^2}\frac{d k_F^{2-d}}{\pi k_F l}\int_0^{1/l}dq \frac{|q|^{d-3}}{\frac{1}{D_0\xi^2} + q^d}
\ee
where we have used that the localization length ${\xi} = \lim_{\omega \to 0}\sqrt{{-\tilde D}(\omega)/i\omega}$.
This expression is obtained by matching the predictions for the conductivity and density response function on the metallic and insulating side of the 
transition \cite{voller,voller1}.
   
%The above expression is already suited for the calculation of critical exponents and the critical disorder
The third term in Eq.(\ref{main}) vanishes   ($\xi \to \infty$) as we approach the MIT. 
 We use this fact to compute the critical disorder, $\lambda =\lambda_c$ and the critical exponent, $s$, related to the vanishing of the conductivity. 
The critical disorder $\lambda_c = \left(\frac{d-2}{\pi^{d-2}d}\right)^{\frac{1}{d-1}}$ is obtained by solving 
$\lambda$ in Eq(\ref{main}) with $\lim_{\omega \to 0} {\tilde D}(\omega) = 0$.  On the metallic side of the 
transition $\lim_{\omega \to 0} {\tilde D}(\omega) = 0$ since  ${\tilde D}(\omega) \propto \sigma(\omega)$ and the conductivity 
$\sigma(0) \propto g_c/L = 0$ for $L \to \infty$. 
% Likewise it is straightforward to show $
%\sigma(0) \propto |\lambda -\lambda_c|^s
%$
%with $s =1$. 
This result agrees with the prediction of Vollhardt and Wolfle \cite{voller1}.  Therefore anomalous diffusion does not affect the behavior of $\lambda_c$ 
close to the transition. 
We note that $\lambda_c = 1/\pi > 0$ for $d \to \infty$ limit. This is consistent with the fact that the MIT in the Cayley tree occurs at finite disorder \cite{MF}.

 The situation is different in the case of the critical exponent $\nu$ related to the divergence of the localization length.
As we approach the transition from the insulator side ($\lambda > \lambda_c$)  the localization length diverges as $\xi \propto |\lambda -\lambda_c|^{-\nu}$. 
The conductivity also vanishes on the insulator side of the transition $\lim_{\omega \to 0} \sigma(\omega) \propto i\omega$ \cite{voller,voller1}.  
Combining the results for the the metallic and insulator sides of the transition it turns out $\lim_{\omega \to 0}{\tilde D}(\omega) \propto i\omega = 0$.
 Using this fact the critical exponent $\nu$, is obtained by 
simply solving Eq.(\ref{main}) for $\xi$ with $\omega \to 0$,
\be
\nu = \frac{1}{d-2} +\frac{1}{2}.
\ee
A few comments are in order,  this expression: a) only agrees with the Vollhardt-Wolfle prediction for $d \sim 2$, b) agrees with numerical calculations for any $d >2$  dimensions \cite{antem2}, 
c)  shows that the upper critical dimension for localization is infinity since $\nu > 1/2$ for any finite $d$.   
This is the most important result of the paper. 

Finally we study the critical dimensional conductance $g_c = \frac{\hbar \tilde D(L)}{L^2\Delta}$. In order to proceed we have to compute ${\tilde D}(L)$. 
 In practical terms this can be carried out by including a lower-cutoff $\sim 1/L$ in the integral over
momentum Eq.(\ref{main}).  For  $\lambda > \lambda_c$, and $\omega \to 0$, ${\tilde D}(\omega) \to 0$ and, 
$
\tilde D (L) = \frac{\Delta}{\pi}\frac{S_d}{(2\pi)^d}
\int_0^{1/L}\frac{|q|^{2d-3}}{\frac{1}{D_0 \xi^2} + q^d}.
$
After performing the integral, and taking the limit $\xi \to \infty$ for a fixed $L$ we obtain, 
\be
\label{g}
g_c = \frac{S_d}{\pi(d-2)(2\pi)^{d}}
\ee
where $S_d$ is the surface of the $d$ sphere. We note that $g_c \ll 1$ for $d \gg 1$.  This result agrees 
with previous predictions based on a self-consistent condition \cite{voller1} or simple one loop perturbation theory \cite{shapiro}.   Thus anomalous diffusion does not affect the value of $g_c$.  
However corrections to the above $g_c$ due to a finite $L$ or $\xi$ will be in
general different from the predictions of  Ref. \cite{voller1,shapiro}.

{\it Level Statistics at the MIT.-}
We study analytically the  
number variance and the two level correlation function (TLCF) at the MIT.   
Our main result is that
the number variance is linear  $\Sigma^2(\ell) \sim  \chi \ell$ with $\chi \approx 1-2/d$. 

We are now interested in the properties of the system precisely at the MIT.
Key in our argument is again the fact that 
diffusion is anomalous at the MIT.
Our starting point is the connected TLCF,  
$
R_{2}(\epsilon_1,\epsilon_2)=\Delta
\langle \rho(\epsilon_1)\rho(\epsilon_2) \rangle \;,
$
($\langle \; \rangle $ denotes averaging over disorder realizations and $\rho$ stands for the spectral density).
In the metallic limit, $g \gg 1$ and for $s \equiv { {\epsilon_1-\epsilon_2} \over \Delta} \gg g$ ,  the TLCF is related to $P(\omega,q_{n_i}) = \frac{1}{-i\omega + 
D_{clas}q^2_{n_i}}$ by,
$
R_{2}(s)= - \frac{\Delta^2}{\pi^2}{\Re}\sum_{n_i}P^2(s\Delta, q_{n_i}),
$
where the sum runs over all momentum eigenstates $q_{n_i}$. This result is semiclassical in the sense that interference corrections represented by maximally crossed 
diagrams are not taken into account.  As disorder increases diffusion becomes slower as a 
consequence of the growing interferences effects. Corrections to the metallic results above are thus expected.  
 
The anomalous diffusion $\langle r^2 \rangle \sim t^{2/d}$ at the MIT is 
reproduced by simply replacing 
the standard diffusion pole $\sim q^2$ in $P(\omega,q)$ by $\sim q^d$.  We thus propose that for $s \gg g_c$ the TLCF
at the transition is given by, 
\be
\label{r2c}
R_{2}(s)= - \frac{1}{\pi^2}{\Re}\sum_{n_i}\frac{1}{(is + g_c|q_{n_i}|^d)^2}
\ee
where 
$|q_{n_i}| = \sqrt{\sum_{i=1}^d {n_i^2}}$. 
In this approach we assume all interference effects are included
in the renormalization of the diffusion coefficient $D_{clas} \to D_0 q^{d-2}$.   Corrections to this result are expected due to to the multifractality of the 
eigenstates. However such corrections cannot modify the scale invariance of $g_c$ at the MIT.

We are now ready to compute the number variance, 
 $\Sigma^2(\ell)=\langle (N_\ell -\langle N_\ell \rangle)^2 \rangle= 2\int_{0}^{\ell}(s - \ell) R_2(s)$  with $N_\ell$ the number of eigenvalues in an interval of length $\ell$ in units of the 
mean level spacing.
Carrying out this integral  and replacing the sum over momenta by an integral,
\be
\Sigma^2(\ell) = \frac{1}{\pi^2}\frac{S_d}{(2\pi)^d}\int_{0}^{\infty}dt |t|^{d-1}\ln \left[\frac{\ell^2}{g_c^2t^{2d}}+1\right].
\ee
Performing the integral and using Eq.(\ref{g}),
\be
 \Sigma^2(\ell) \approx \chi \ell~~~~~\chi = 1-2/d ~~~~~~\ell \gg g_c
\ee
 %We note that  the range of validity of the perturbative expression Eq. \cite{r2c} is precisely the one 
%that controls the asymptotic behavior of the number variance. 
A linear number variance with $\chi < 1$ is considered a signature of a MIT.  
The origin of this linear behavior was predicted heuristically
\cite{chi} by using OPT and making the plausible approximation
that eigenvalues interact only if their separation is
smaller than $g_c$.  The value of the slope was later estimated to be $\chi = {{d-D_2} \over {2d}} $ \cite{CK}. We do 
not fully understand the relation between this result and ours $\chi = 1-2/d$. However we note 
our expression for $\chi$ reproduces correctly the limits $d \sim 2$ and $d \to \infty$. 
For $d \geq 3$ our prediction is around $10-15\%$ off the numerical value \cite{antem2}, the  predictions of Ref. \cite{CK} fails for $d >3$.  
%Another reason could be the numerical value of the slope gets corrections from the region $s \leq g_c$ as in the case of  power-law 
%random banded models \cite{ever}.

We now turn to the discussion of the TLCF in the region $s \sim g_c$. We aim to examine the
heuristic arguments of Ref.\cite{chi} where it was suggested that for $s > g_c$ there must be a sharp suppression of the spectral correlations.
If localization corrections are negligible the use of the supersymmetry method permits an explicit evaluation of the nonperturbative part 
of the TLCF \cite{andre}.
%In the region $s \ll g$ \cite{KM} the effect of a finite $g$ is just a small correction to the universal results of random matrix theory.
%For larger energy separations \cite{andre} it was found that the non perturbative part of the TLCF 
$
R_{2}^{NP}(s) \propto D^2(s,g)$
where   
$D(s,g)= \prod_{n_i \neq 0}(1+{s^{2} \over {q_{n_i}^{2d}g^2}})^{-1}$  is the spectral determinant associated to the classical diffusion operator $P(\omega,q)$. 
It is plausible to expect that such expression for $R_2^{NP}$ can still be used at the MIT provided that the spectral determinant is modified to take into account the anomalous diffusion predicted by the OPT, namely,  $|q_{n_i}| = \sqrt{\sum_{i=1}^d {n_i^2}}$ and $g \to g_c$.  This is again a semiclassical approximation, we suggest that all the quantum interference effects at the MIT are included by an appropriate redefinition of the classical spectral determinant.  
 The spectral determinant $D(s,g_c)$ can then be estimated analytically by exponentiation of the product,  replacing sums by integrals and using Eq.(\ref{g}).
 The final result is simply, 
\be 
R_2^{NP}(s) \propto e^{- 2 \pi^2 s (d-2)/d}
\ee
We note, a) an exponential decay with a similar prefactor has been observed in numerical calculation \cite{antem2}, b) the sharp decay 
of the TLCF does not occur for $s \sim g_c \ll 1$ but rather for $s \sim 2d/(d-2)\pi \gg g_c$, c) the conformal symmetry predicted by the OPT at the MIT 
is only consistent with an exponential or a power-law decay of the TLCF, b) the power-law decay of $R_2(s)$ observed in Ref. \cite{aronov} is related to how the system approach the transition rather than to the transition itself. 
 
%However  
%the exponential contribution seems to be the one responsible for critical features such as a linear number variance.

{\it Limits of applicability.-}
According to numerical and heuristic arguments \cite{huck} it is expected that $P(q,\omega)$ will depend on the multifractal dimension $D_2$ rather than $d$ for times and distances much shorter than the Heisenberg time and the system size respectively.
The proposed renormalization of the diffusion coefficient ($D(q) \propto q^{d-2}$) is in principle restricted to the region $q, \omega\to 0$. However we note this region is the only relevant in the calculation of the critical exponents. In the evaluation of the number variance $\Sigma^(\ell)$ other momentum regions may also be relevant.  Therefore our expression of the slope as a function of $d$ may get corrections depending on the multifractal dimension $D_2$ \cite{CK}.  These corrections point to the limit of applicability of our approach. The OPT is 
based on the scaling of the moments not on the distribution function itself.  However multifractality is a property of the distribution function and consequently beyond the reach of the OPT formalism.

In conclusion, we have put forward an analytical approach to the metal-insulator transition based on the anomalous diffusion 
predicted by the OPT at the critical point.  With this simple input we have shown that the upper critical dimension is infinity and found explicit expressions for the critical exponents and critical disorder as a function of the spatial dimensionality.  Moreover we have shown that the number variance is asymptotically linear with a slope which is a simple increasing function of the spatial dimensionality.  All our analytical predictions are in fair agreement with numerical simulations. \\

I  thank Denis Basko and Emilio Cuevas for illuminating conversations. 
% References %%%%%%%%%%%%%%%%%%%%%%%%%%%%%%%%%%%%%%%%%%%%%%%%%%%%%%%%%%%%%%%%%%%%%

\end{document}